\documentclass[referee]{raa}            

\usepackage{graphicx,times}             
\usepackage{natbib}
\usepackage{amssymb,amsmath}
\bibpunct{(}{)}{;}{a}{}{,}
\usepackage{hyperref}
\usepackage{color,colortbl}
\hypersetup{pdftitle = The title of my PDF, pdfauthor = My name,
pdfsubject= The subject, pdfkeywords = keyword1 keyword2 keyword3}
\hypersetup{colorlinks = true, linkcolor = green, citecolor =
blue}
\definecolor{lightRed}{RGB}{230,170,150}

\begin{document}

   \title{The period-luminosity relation for Cepheids derived from
    multiphase temperature measurements and Cepheids kinematics based on GAIA
    DR2 data
}

   \volnopage{Vol.0 (20xx) No.0, 000--000}      
   \setcounter{page}{1}          

   \author{Y. A. Lazovik\inst{1,2}
   \and A. S. Rastorguev\inst{1,2}
   \and M. V. Zabolotskikh\inst{2}
   \and N. A. Gorynya\inst{2,3}
   }

   \institute{Lomonosov Moscow State University, Faculty of Physics,
1 Leninskie Gory, bldg.2, Moscow, 119991, Russia;
\\{\it E-mail: yaroslav.lazovik@gmail.com}\\
\and Lomonosov Moscow State University, Sternberg Astronomical
Institute, 13 Universitetskii prospect, Moscow, 119992, Russia;
\\ \and Instutute of Astronomy RAS, 48 Pyatnitskaya str., Moscow,
119017, Russia\\
\vs \no
   {\small Received~~20xx month day; accepted~~20xx~~month day}}

\abstract{ Calibration of the period-luminosity relation (PLR) for
Cepheids has always been one of the biggest goals of stellar
astronomy. Among a considerable number of different approaches,
the Baade-Becker-Wesselink (BBW) method stands in the foreground
as one of the most universal and precise methods. We present a new
realization of the BBW method which is considered to be the
generalization of surface brightness (\citealt{Barnes+Evans+1976})
and \cite{Balona+1977} approaches first proposed by
\cite{Rastorguev+Dambis+2010} and described in \cite{Paper1}. One
of the main features of this method is using measured effective
temperature variations to determine the main parameters of
Cepheid, such as distance, radius, luminosity, colour excess,
intrinsic colour. We apply this method to 45 Cepheids of Northern
sky, for which multiphase temperature data are available. We take
into account the effect of shock waves, whose presence in stellar
atmosphere distorts the observational data and the calibrations
used in this work. Within $0.0-0.87$ phase interval we derived PL
relation $<M_V>_I=-(2.67 \pm 0.17)\cdot logP - (1.58 \pm 0.16)$.
It was used to calculate the distances, rotation curve and
kinematical parameters of the sample of 435 Cepheids with GAIA DR2
proper motions.  }

   \authorrunning{Y. Lazovik, A. Rastorguev, M. Zabolotskikh, N. Gorynya }            
   \titlerunning{New period-luminosity relation for Cepheids}  

   \maketitle
\section{Introduction}
\label{sect:intro} In the modern astronomy Cepheids play a
 particularly important role. PL relation, which was first discovered in
1912, made these objects one of the most reliable standard candles
in the context of the extragalactic distance scale calibration.
Nowadays obtaining precise PL relation remains one of the priority
astronomical goals.

Many different approaches have been proposed in order to solve
this task, among which several deserve special attention. One of
the most commonly used methods to determine distance is the method
of trigonometric parallax, which is inextricably linked to the
GAIA mission (\citealt{Gaia}).  However, the distances derived
with this method are fraught with significant level of uncertainty
and systematic errors (\citealt{Groenewegen+2018}). In comparison
with trigonometric parallaxes, the distances obtained for Cepheids
in open clusters are more accurate, but limited number of such
objects makes PL relation based on Cepheids in open clusters less
reliable. The Baade-Becker-Wesselink (BBW; \citealt{Baade+1926};
\citealt{Becker+1940}; \citealt{Wesselink+1946}) method is thought
to be quite effective and universal and comparable to the
approaches mentioned above. One of the most well-known
implementations of this method is the surface-brightness technique
(\citealt{Barnes+Evans+1976}); including infrared range
(\citealt{Barnes+2005}, hereafter \textbf{IRSB}); nevertheless
it's not the only one.

In our research we focus on another modification of the BBW
method, namely maximum-likelihood technique (i.e. light curve
modelling), first proposed by \cite{Balona+1977} and recently
generalized and refined by \cite{Rastorguev+Dambis+2010},
\cite{Rastorguev+2013} (hereafter \textbf{RD} version). This
modification allows one to determine all the main parameters of
Cepheid including mean radius, luminosity, color excess and normal
color, as well as apparent distance modulus. Now we present first
results of the study that is based on our new approach which uses
additionally published multiphase effective temperature
measurements (see, for example, catalog \cite{Luck+2018}). Here we
estimate the luminosities of Cepheids and derive new calibration
of the PL relation as described in \citet{Paper1}.

\section{Observational data and samples of Cepheids}
\label{sect:data}

In this study we use photoelectric and CCD photometry of classical
Cepheids from \cite{Berdnikov}, very accurate radial-velocity
measurements published by
\cite{Gorynya+1992,Gorynya+1996,Gorynya+1998,Gorynya+2002} and
effective temperature data from \cite{Luck+2018} catalog.

We have selected the data sets according to quality and
completeness, but also to ensure that photometric and
spectroscopic observational data are as synchronous as possible to
prevent any systematic errors in the computed radius value and
other parameters owing to evolutionary period changes resulting in
phase shifts between the light, color and radial-velocity curves.

The above \textbf{RD} calculation algorithm (\citealt{Paper1}) has
been conducted for 45 Cepheids. These Cepheids were divided into
three samples according to the expected accuracy of the final
results. Several factors were taken into account, including the
quality of the observational data, its uniformity on the phase
curve, the presence of a component (for binary/multiple objects we
carried out additional calculations in order to extract the
pulsation velocity curve from the radial velocity data and to
subtract hot component's radiation from the photometric data), the
type of pulsation (overtone or fundamental), the expected position
in the instability strip (in the center or on the edge) and
others. The first sample, for which we expect the most precise
results, contains 23 Cepheids, the second one includes 10 objects
and the rest 12 Cepheids belong to the third sample. The list of
Cepheids and sample memberships are given in Table ~\ref{tab1}.

\begin{table}
\bc
\begin{minipage}[]{100mm}
\caption[]{Parameters of 45 Cepheids with [0, 0.87] phase interval\label{tab1}}\end{minipage}
\setlength{\tabcolsep}{1pt}
\small
\renewcommand{\tabcolsep}{0.3cm}
 \begin{tabular}{ccccccccccccc}
  \hline\noalign{\smallskip}
Name&Sample&Binary&Fundamental period&$E(B-V)$&$<R>$&$<{M_v>_I}$&$(m - M)_0$\\
&&&($days$)&($mag$)&($R_\odot$)&($mag$)&($mag$)\\
  \hline\noalign{\smallskip}
AW Per&3&Yes&6.463&0.58&38.8 $\pm$ 1.6&-3.31 $\pm$ 0.16&8.94 $\pm$ 0.20\\
BB Her&1&No&7.508&0.40&54.5 $\pm$ 2.7&-3.78 $\pm$ 0.07&12.54 $\pm$ 0.11\\
BE Mon&3&No&3.811&0.57&44.1 $\pm$ 12.8&-3.55 $\pm$ 0.31&12.24 $\pm$ 0.34\\
BG Lac&2&No&5.332&0.29&43.8 $\pm$ 2.0&-3.34 $\pm$ 0.11&11.28 $\pm$ 0.13\\
CD Cyg&1&No&17.074&0.58&103.1 $\pm$ 1.9&-4.93 $\pm$ 0.06&11.95 $\pm$ 0.14\\
CF Cas&1&No&4.875&0.54&45.8 $\pm$ 1.2&-3.41 $\pm$ 0.06&12.78 $\pm$ 0.13\\
CV Mon&3&No&5.379&0.69&52.4 $\pm$ 2.4&-3.82 $\pm$ 0.16&11.84 $\pm$ 0.21\\
Delta Cep&1&Yes&5.366&0.09&46.6 $\pm$ 2.5&-3.65 $\pm$ 0.08&7.30 $\pm$ 0.08\\
DL Cas&3&Yes&11.268&0.65&87.0 $\pm$ 4.8&-4.82 $\pm$ 0.16&11.70 $\pm$ 0.21\\
DT Cyg&3&No&3.520&0.04&42.3 $\pm$ 4.7&-3.63 $\pm$ 0.18&9.28 $\pm$ 0.19\\
Eta Aql&1&No&7.177&0.17&60.6 $\pm$ 2.9&-4.09 $\pm$ 0.07&7.43 $\pm$ 0.08\\
FF Aql&3&Yes&6.297&0.27&53.7 $\pm$ 7.5&-4.16 $\pm$ 0.20&8.64 $\pm$ 0.21\\
FM Aql&1&No&6.114&0.69&52.6 $\pm$ 1.6&-3.81 $\pm$ 0.06&9.81 $\pm$ 0.15\\
FN Aql&1&No&9.482&0.48&62.2 $\pm$ 1.3&-3.87 $\pm$ 0.06&10.67 $\pm$ 0.12\\
RS Ori&3&No&10.658&0.37&72.7 $\pm$ 4.2&-4.60 $\pm$ 0.16&11.78 $\pm$ 0.18\\
RT Aur&1&No&3.728&0.06&36.7 $\pm$ 1.5&-3.22 $\pm$ 0.07&8.47 $\pm$ 0.07\\
RX Aur&1&No&11.624&0.34&71.8 $\pm$ 2.2&-4.48 $\pm$ 0.06&11.01 $\pm$ 0.10\\
RX Cam&2&Yes&7.912&0.55&53.0 $\pm$ 3.2&-3.78 $\pm$ 0.12&9.65 $\pm$ 0.16\\
S Sge&1&Yes&8.382&0.17&52.5 $\pm$ 1.2&-3.78 $\pm$ 0.06&8.82 $\pm$ 0.07\\
S Vul&1&No&68.438&1.15&250.8 $\pm$ 7.6&-6.93 $\pm$ 0.12&12.10 $\pm$ 0.26\\
SS Sct&2&No&3.671&0.39&36.0 $\pm$ 0.9&-3.13 $\pm$ 0.12&10.06 $\pm$ 0.14\\
SU Cyg&3&Yes&5.417&0.09&58.1 $\pm$ 6.1&-4.30 $\pm$ 0.19&10.88 $\pm$ 0.19\\
SV Mon&1&No&15.235&0.30&95.8 $\pm$ 1.6&-4.71 $\pm$ 0.08&11.99 $\pm$ 0.10\\
SV Vul&1&No&44.969&0.62&192.2 $\pm$ 3.1&-6.08 $\pm$ 0.07&11.42 $\pm$ 0.14\\
T Mon&1&Yes&27.033&0.31&120.1 $\pm$ 1.6&-4.96 $\pm$ 0.06&10.10 $\pm$ 0.09\\
T Vul&1&Yes&4.435&0.07&39.9 $\pm$ 1.4&-3.33 $\pm$ 0.06&8.88 $\pm$ 0.07\\
TT Aql&1&No&13.755&0.59&86.8 $\pm$ 1.8&-4.58 $\pm$ 0.07&9.76 $\pm$ 0.14\\
U Aql&3&Yes&7.024&0.44&47.7 $\pm$ 1.9&-3.66 $\pm$ 0.18&8.62 $\pm$ 0.20\\
U Sgr&1&No&6.745&0.46&50.6 $\pm$ 2.2&-3.70 $\pm$ 0.07&8.89 $\pm$ 0.12\\
U Vul&2&Yes&7.990&0.72&43.3 $\pm$ 1.2&-3.47 $\pm$ 0.06&8.21 $\pm$ 0.16\\
V500 Sco&2&No&9.317&0.62&62.1 $\pm$ 3.8&-4.05 $\pm$ 0.13&10.75 $\pm$ 0.18\\
VX Per&2&No&10.885&0.53&79.8 $\pm$ 3.3&-4.61 $\pm$ 0.11&12.15 $\pm$ 0.16\\
W Gem&2&No&7.914&0.30&49.4 $\pm$ 1.7&-3.71 $\pm$ 0.11&9.65 $\pm$ 0.13\\
W Sgr&1&Yes&7.595&0.13&50.6 $\pm$ 1.5&-3.74 $\pm$ 0.07&7.99 $\pm$ 0.07\\
WZ Sgr&1&No&21.850&0.59&114.2 $\pm$ 1.9&-4.92 $\pm$ 0.11&11.01 $\pm$ 0.16\\
X Cyg&1&No&16.386&0.34&94.0 $\pm$ 2.4&-4.54 $\pm$ 0.08&9.80 $\pm$ 0.11\\
X Pup&3&No&25.965&0.53&107.8 $\pm$ 2.5&-5.15 $\pm$ 0.16&11.93 $\pm$ 0.19\\
X Vul&1&No&6.320&0.85&50.7 $\pm$ 1.8&-3.76 $\pm$ 0.06&9.78 $\pm$ 0.18\\
XX Sgr&2&No&6.424&0.58&52.9 $\pm$ 2.0&-3.97 $\pm$ 0.11&10.93 $\pm$ 0.16\\
Y Lac&3&No&6.090&0.15&52.5 $\pm$ 2.6&-3.94 $\pm$ 0.16&12.58 $\pm$ 0.16\\
Y Oph&2&No&17.128&0.78&105.7 $\pm$ 3.1&-5.33 $\pm$ 0.10&8.92 $\pm$ 0.19\\
Y Sgr&2&No&5.773&0.23&49.7 $\pm$ 1.1&-3.70 $\pm$ 0.06&8.67 $\pm$ 0.07\\
YZ Sgr&1&No&9.554&0.36&55.4 $\pm$ 1.5&-3.79 $\pm$ 0.06&9.96 $\pm$ 0.09\\
Z Lac&1&Yes&10.886&0.48&70.0 $\pm$ 1.1&-4.28 $\pm$ 0.07&11.10 $\pm$ 0.12\\
Zet Gem&3&No&10.150&0.07&67.2 $\pm$ 3.2&-4.06 $\pm$ 0.16&7.69 $\pm$ 0.16\\
  \noalign{\smallskip}\hline
\end{tabular}
\ec
\end{table}

\section{Results and discussion}
\label{sect:results} Obtaining PR relation allows one to identify
the overtone Cepheids. These objects are shifted towards lower
period values on the PR diagram. Among the Cepheids studied in the
present work seven objects have been suspected to be overtone
pulsators based their positions on period-radius and
period-luminosity diagrams, namely BE Mon, DL Cas, DT Cyg, FF Aql,
RS Ori, SU Cyg and Y Lac. The periods of these stars were shifted
by a $\Delta \log P^d\approx +0.15$ to fit the fundamental period.
All suspected overtone pulsators were included to the third
sample, meaning the parameters estimated for these Cepheids are
less reliable.
\begin{figure}[h]
  \begin{minipage}[t]{0.495\linewidth}
  \centering
   \includegraphics[width=80mm]{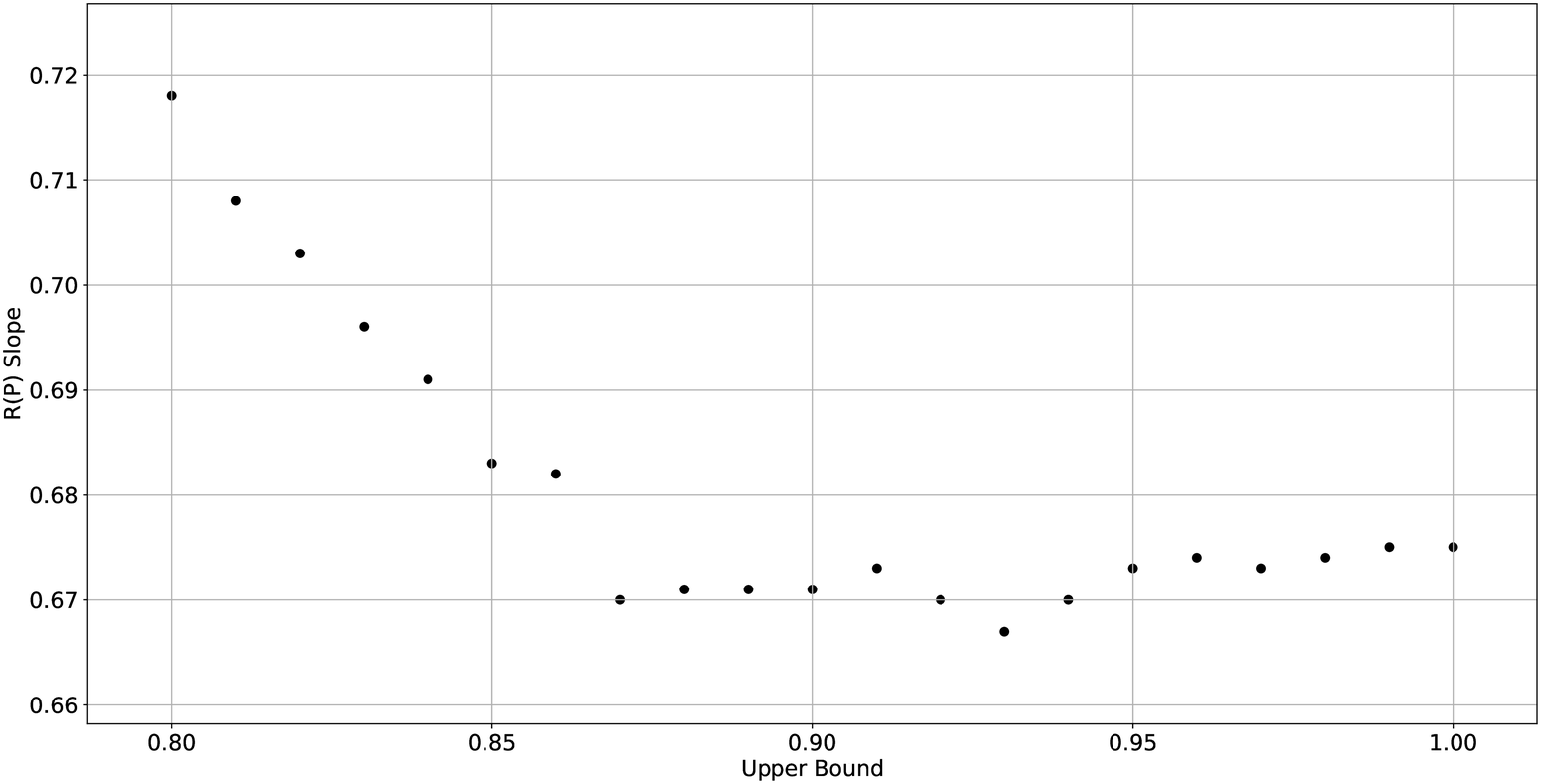}
   \caption{{\small Slope of the PR relations versus upper phase constraint}}
   \label{fig:fig1}
  \end{minipage}%
  \begin{minipage}[t]{0.495\textwidth}
  \centering
   \includegraphics[width=80mm]{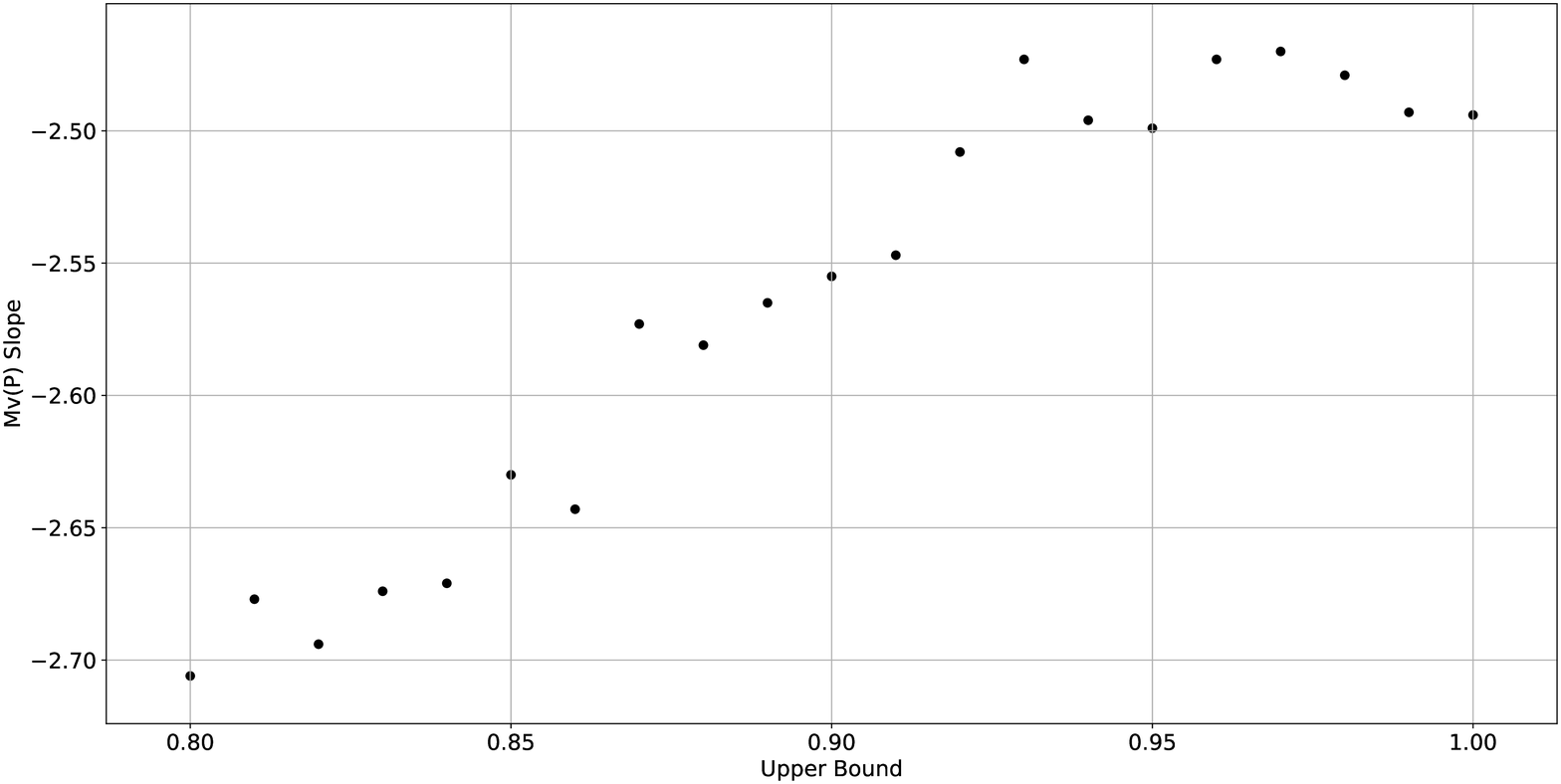}
  \caption{{\small  Slope of PL relations versus upper phase constraint}}
  \label{fig:fig2}
  \end{minipage}%
\end{figure}
\begin{figure}[h]
  \centering
   \includegraphics[width=120mm]{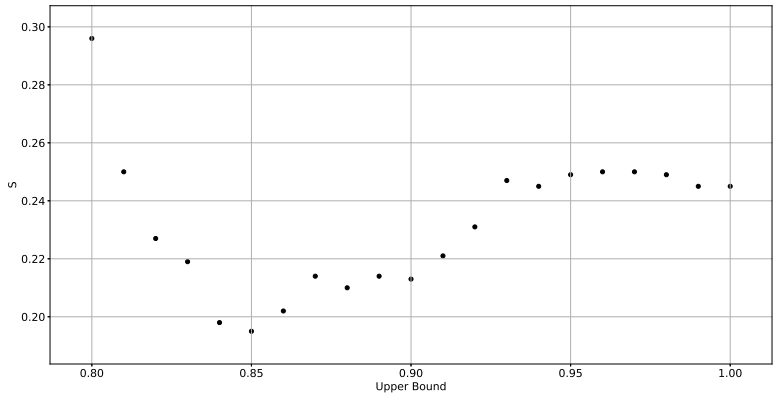}
   \caption{{\small Standard deviation from linear PL
    relation versus upper phase constraint}}
   \label{fig:fig3}
\end{figure}

As is known, at the moments of time, corresponding to the upper
phases of pulsation (near $\varphi \approx 1.0$), shock waves can
arise in stellar atmosphere. These shock waves change the physical
conditions in the environment and produce asymmetric line profiles
in the spectrum which leads to the mismatch between the
observational data and the calibrations. Moreover, the propagation
of shock waves may be reflected in the change of the projection
factor. We therefore need to constrain our data to avoid the
distorted observations. Some authors (for example,
\citealt{Storm2}) exclude phase region [0.8, 1.0] from
consideration. We decided to repeat our algorithm for the first
sample Cepheids with different upper phase limits and find out
which constraint leads to the smallest scatter of individual
Cepheids relative to linear PL relation. As shown in Fig.
~\ref{fig:fig1} and ~\ref{fig:fig2}, the choice of constraint
decisively affects the obtained PR and PL relations. From our
point of view, the optimal restriction for the phase interval is
[0.00, 0.87]. This choice doesn't change the slope of PR relation
compared to the case without restriction (Fig. ~\ref{fig:fig1})
and reduces the scatter relative to the linear PL relation (Fig.
~\ref{fig:fig3}). It is worth noting that the spread of the third
sample Cepheids relative to the linear PL dependence has become
slightly larger, therefore, we recommend to use the expressions
obtained for the combination of the first and the second samples.
\begin{figure}[h]
  \begin{minipage}[t]{0.495\linewidth}
  \centering
   \includegraphics[width=80mm]{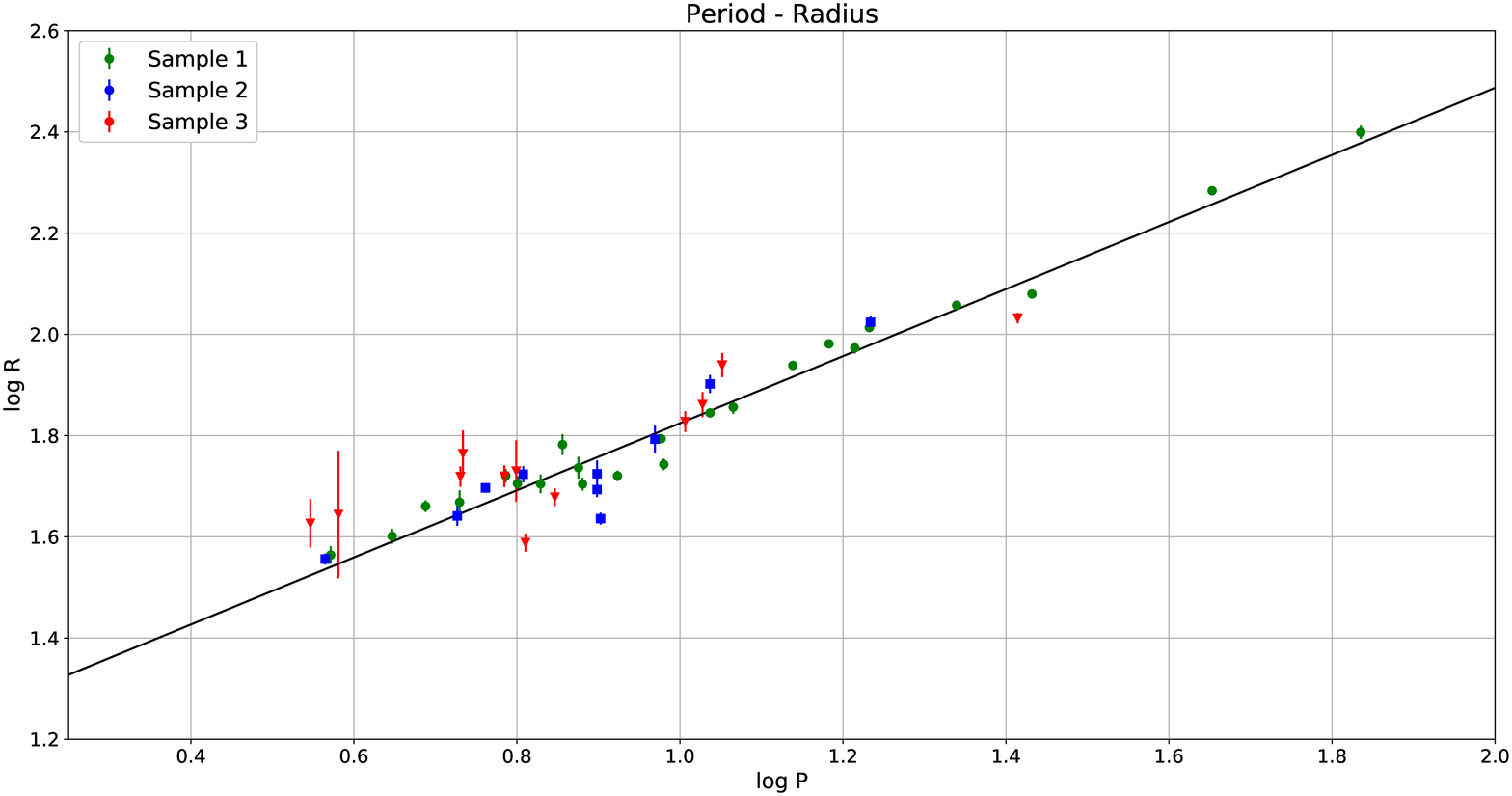}
   \caption{{\small Period-Radius diagram, corresponding to the optimal phase interval  } }
   \label{fig:fig4}
  \end{minipage}%
  \begin{minipage}[t]{0.495\textwidth}
  \centering
   \includegraphics[width=80mm]{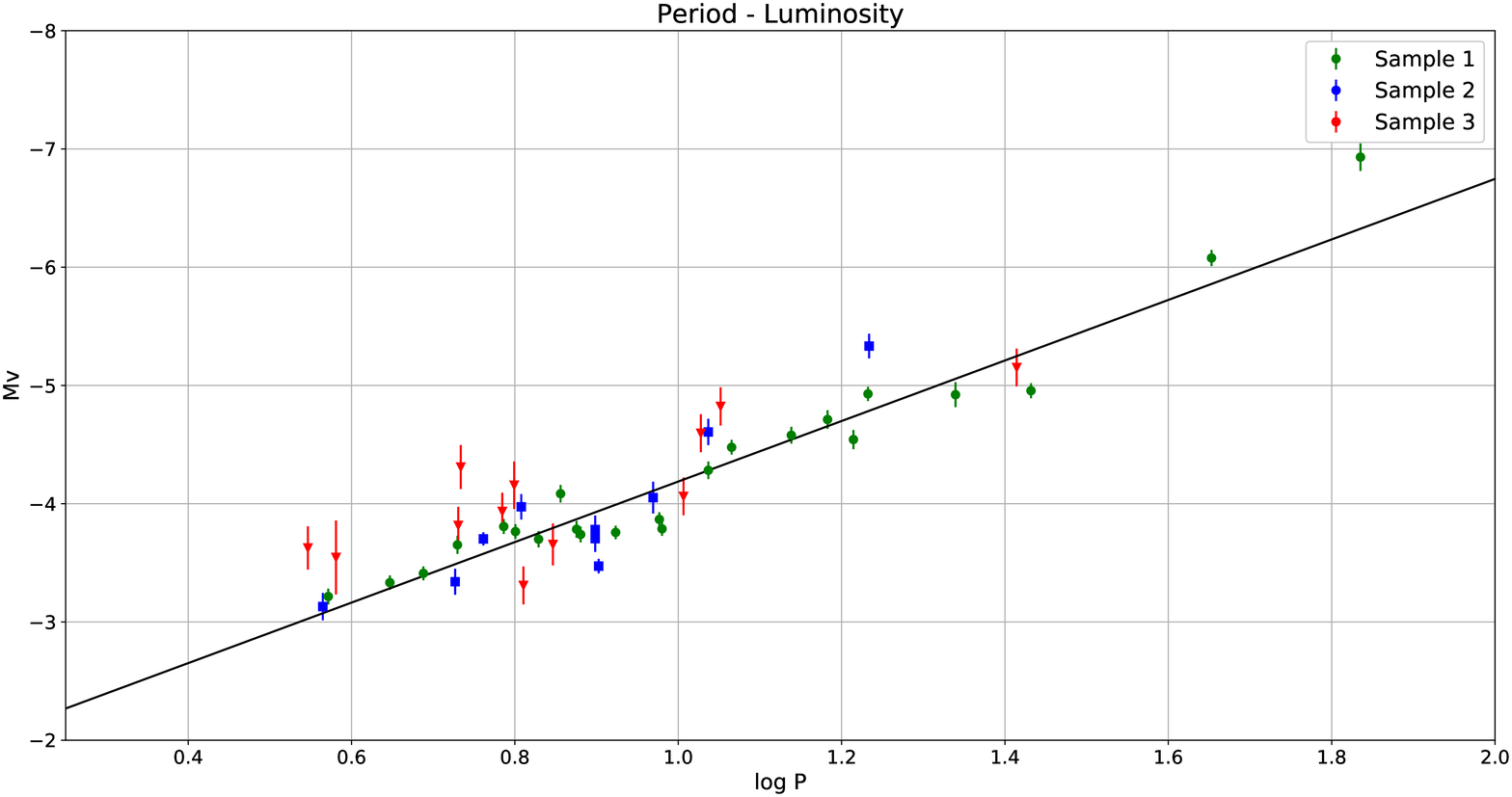}
  \caption{{\small  Period-Luminosity diagram, corresponding to the optimal phase interval }}
  \label{fig:fig5}
  \end{minipage}%
\end{figure}

PR and PL relations are shown in Fig. ~\ref{fig:fig4} and
~\ref{fig:fig5}, respectively. Figures demonstrate that there's a
strong correlation between the quality of the sample (the order
number in the Table ~\ref{tab1}) and the scatter relative to
linear relationship. Leaving the first two samples and adopting
the optimal phase constraint, we obtained the following PR
relation: $logR = (0.67 \pm 0.02)\cdot logP + (1.15 \pm 0.03)$, in
good agreement with other works (for example,
\citealt{Sachkov+1998}).

\begin{table}
\begin{center}
\caption[]{ The period-luminosity relations in the form $M = a
\cdot  log(P)  + b$\label{tab2}}\

\setlength{\tabcolsep}{1pt} \small
\renewcommand{\tabcolsep}{0.5cm}
 \begin{tabular}{cccccccccccccccccccccc}
  \hline\noalign{\smallskip}
References &  a      & b & Method                    \\
  \hline\noalign{\smallskip}
\cite{Groenewegen+2018}&-2.24 $\pm$ 0.14&-1.84 $\pm$ 0.12&GAIA DR2 trigonometric parallax \\
\cite{Benedict}&-2.43 $\pm$ 0.12& -1.62 $\pm$ 0.02&HST trigonometric parallax  \\
\rowcolor{lightRed}
Present work [0.00; 1.00]&-2.45 $\pm$ 0.15 & -1.79 $\pm$ 0.14&RD \\
\rowcolor{lightRed}
Present work [0.00; 0.87]&-2.60 $\pm$ 0.17 & -1.58 $\pm$ 0.16&RD \\
\cite{Gieren}&-2.62 $\pm$ 0.10 & -1.37 $\pm$ 0.04&IRSB \\
\cite{Storm2}&-2.67 $\pm$ 0.10 & -1.29 $\pm$ 0.03&IRSB \\
\cite{Fouque}&-2.68 $\pm$ 0.09 & -1.28 $\pm$ 0.03&IRSB \\
\cite{Molinaro}&-2.78 $\pm$ 0.11 & -1.42 $\pm$ 0.11&IRSB \\
\cite{Turner}&-2.78 $\pm$ 0.12 & -1.29 $\pm$ 0.10&Membership in open clusters\\
  \noalign{\smallskip}\hline
\end{tabular}
\end{center}
\end{table}

The coefficients of our PL relations and the PL relations from the
literature are listed in Table ~\ref{tab2}. It's clearly seen that
in terms of the slope our relations fill the gap between the
relations obtained with trigonometric parallaxes and the relations
obtained with \textbf{IRSB} version of the BBW method, which seems
to be quite satisfactory result. The deviation of our relation
from the relation \cite{Groenewegen+2018} is small in the region
of the short period Cepheids and becomes larger as the period
increases, which is also expected as the accuracy of trigonometric
parallaxes becomes lower for the far located stars, among which
there are statistically more bright ones (the latter is explained
by selection effects).

\begin{figure}[h]
  \centering
   \includegraphics[width=140mm]{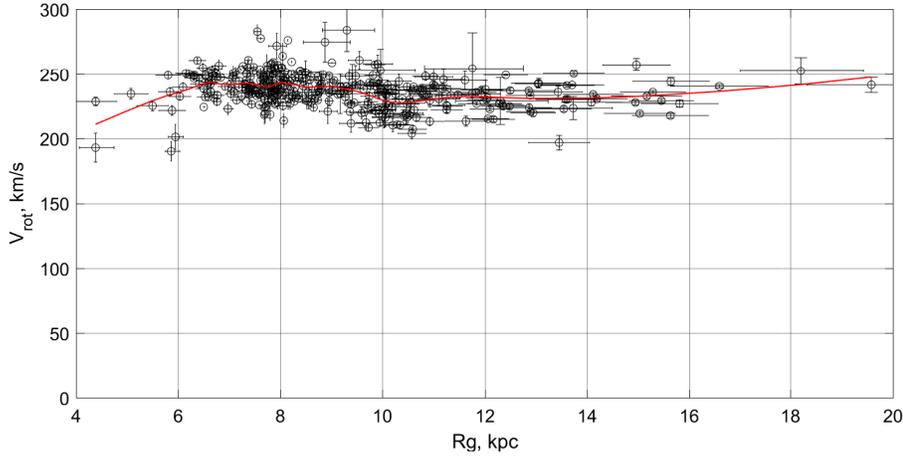}
   \caption{{\small Rotation curve of the Galactic disk built
   with statistical-parallax technique using obtained PL relation}}
   \label{fig:fig6}
\end{figure}

To verify our result from completely different kinematical
approach we checked the applicability of our \textbf{RD} method
with statistical-parallax technique (\citealt{Rastorguev+2017}).
We combined flux averaged V magnitudes and the radial velocities
from \cite{Melnik+2015}; averaged values of the color excess and
the radial velocities from the DDO database (\citealt{Fernie});
accurate radial velocities and proper motions from GAIA DR2
catalogue (\citealt{Gaia}) and our new PL relation to derive the
distances for a large sample of Cepheids and to study their
kinematics. The final sample with uniform photometric distances
and spatial velocities consists of 435 stars. The calculation of
kinematic parameters of the velocity field, which includes
differential rotation and small perturbations from four-arm spiral
density wave, was carried out by the statistical parallax method,
described in all details in \cite{Rastorguev+2017}.

The distance to the center of the Galaxy was taken to be $R_0 =
8.2$ kpc. As for galactic masers, the best model of the velocity
field presumed radial velocity dispersion independent on the
galactocentric distance. The kinematic parameters and their
statistical errors are listed in Table ~\ref{tab3}. They include
local velocity of the sample $(U_0, V_0, W_0)$ relative to the
Sun, radial and vertical velocity dispersion ($\sigma U_0, \sigma
W_0$), amplitudes of the velocity perturbations ($f_R, f_\Theta$),
the Solar phase angle $\chi _0$ in the spiral pattern and pitch
angle $i$, and also the distance scale factor $P = r_ {adopted} /
r_ {true}$. The values of the kinematic parameters are in general
agreement with the data on other young objects: masers, OB-stars
and open clusters (\citealt{Rastorguev+2017, Bobylev+2018a,
Bobylev+2018b, Bobylev+2019a, Bobylev+2019b}). The value of $P$
determines the ratio of the adopted distance and the ``true''
distance derived with statistical-parallax technique. Within
calculated error the distance scale factor is equal to unity which
confirms the quality of our PL relation and the absence of some
systematic error in its zero-point. Fig.~\ref{fig:fig6} shows the
rotation curve of the Galaxy in the galactocentric distance
interval from 4 to 20 kpc. The Solar velocity is about $239 \pm 4$
km/s.

\begin{table}
\begin{center}
\caption[]{ Kinematic parameters obtained with
statistical-parallax technique for the final sample of 435
Cepheids (see \citealt{Rastorguev+2017} for details
)
\label{tab3}}\

\setlength{\tabcolsep}{1pt}
\small
\renewcommand{\tabcolsep}{0.5cm}
 \begin{tabular}{cccccccccccccccccccccc}
  \hline\noalign{\smallskip}
$U_0$&$V_0$&$W_0$&$\sigma U_0$&$\sigma W_0$&$\textit{\textbf{P}}$  \\
($km/s$)&($km/s$)&($km/s$)&($km/s$)&($km/s$)&\\
-10.0 $\pm$ 1.2&-11.8 $\pm$ 0.8&-7.3 $\pm$ 1.1&15.8 $\pm$ 0.6&8.3 $\pm$ 1.0&\textbf{1.01 $\pm$ 0.02}&\\
  \noalign{\smallskip}\hline
  $f_R$&$f_ \Theta$&$\chi_0$&$i$&$\omega _0$&$\omega'_0$ \\
  ($km/s$)&($km/s$)&($deg.$)&($deg.$)&($km/s/kpc$)&($km/s/kpc^2$)\\
  -1.9 $\pm$ 0.8&1.2 $\pm$ 1.5&152 $\pm$ 25&-13.6 $\pm$ 1.7&29.20 $\pm$ 0.40&-3.92 $\pm$ 0.05 \\
  \hline\noalign{\smallskip}
  $\omega''_0$ & $\omega'''_0$\\
  ($km/s/kpc^3$)&($km/s/kpc^4$) \\
  0.74 $\pm$ 0.03&-0.08 $\pm$ 0.01 \\
  \hline\noalign{\smallskip}
\end{tabular}
\end{center}
\end{table}
We are thus convinced that new \textbf{RD} version of BBW method
gives reliable results and in the future it certainly has the
potential to become one of the main tools for a distance scales
calibration.

\begin{acknowledgements}

We thank Russian Foundation for Basic Research (grants nos.
18-02-00890 and 19-02-00611) for partial financial support.
\\

\end{acknowledgements}

\bibliographystyle{raa}
\bibliography{bibtex}

\end{document}